\documentclass[preprint,12pt]{elsarticle}

\usepackage{amssymb}
\usepackage{amsmath}
\usepackage{url}

\journal{xxxx}

\begin{document}

\begin{frontmatter}



\title{The influence of the random numbers quality on the results in stochastic simulations and machine learning}

\author[inst1]{Antunes Benjamin}

\affiliation[inst1]{organization={University of Perpignan via domitia},
            city={Perpignan},
            postcode={66100}, 
            country={France}}

\begin{abstract}
Pseudorandom number generators (PRNGs) are ubiquitous in stochastic simulations and machine learning (ML), where they drive sampling, parameter initialization, regularization, and data shuffling. While widely used, the potential impact of PRNG statistical quality on computational results remains underexplored. In this study, we investigate whether differences in PRNG quality, as measured by standard statistical test suites, can influence outcomes in representative stochastic applications. Seven PRNGs were evaluated, ranging from low-quality linear congruential generators (LCGs) with known statistical deficiencies to high-quality generators such as Mersenne Twister, PCG, and Philox. We applied these PRNGs to four distinct tasks: an epidemiological agent-based model (ABM), two independent from-scratch MNIST classification implementations (Python/NumPy and C++), and a reinforcement learning (RL) CartPole environment. Each experiment was repeated 30 times per generator using fixed seeds to ensure reproducibility, and outputs were compared using appropriate statistical analyses. Results show that very poor statistical quality, as in the “bad” LCG failing 125 TestU01 Crush tests, produces significant deviations in ABM epidemic dynamics, reduces MNIST classification accuracy, and severely degrades RL performance. In contrast, mid- and good-quality LCGs—despite failing a limited number of Crush or BigCrush tests—performed comparably to top-tier PRNGs in most tasks, with the RL experiment being the primary exception where performance scaled with statistical quality. Our findings indicate that, once a generator meets a sufficient statistical robustness threshold, its family or design has negligible impact on outcomes for most workloads, allowing selection to be guided by performance and implementation considerations. However, the use of low-quality PRNGs in sensitive stochastic computations can introduce substantial and systematic errors.
\end{abstract}


\end{frontmatter}



\section{Introduction}
\label{sec:Introduction}

Randomness is a cornerstone of modern computational science, forming the basis for algorithms in both stochastic simulations and ML. In practice, the majority of applications use PRNGs rather than true random number sources. PRNGs are deterministic algorithms that generate sequences of numbers designed to mimic the statistical properties of truly random sequences, enabling reproducibility, efficiency, and scalability. The quality of the generated numbers---determined by statistical properties such as uniformity, independence, and period length---can have a direct impact on the stability, accuracy, and reproducibility of results.

\par
The usage of PRNGs in ML is widespread. Many core algorithms rely on randomness as an integral part of their functioning. A notable example is stochastic gradient descent (SGD), a cornerstone optimization algorithm for training models in ML and deep learning. SGD operates by using a single or small batch of training samples to compute the gradient and update parameters, rather than processing the entire dataset at once. Lu et al.~\cite{lu2022general} demonstrated that using a quasi-Monte Carlo method can accelerate convergence rates for learning with data augmentation, also employing a fixed scan order to improve efficiency.

\par
Randomness also underpins key regularization methods. Dropout, for example, combats overfitting by randomly omitting neurons and their connections during training, improving generalization on unseen data. Stochastic depth, another regularization technique, addresses challenges in deep convolutional networks such as vanishing gradients and long training times by randomly removing layers during each batch and connecting the remaining layers via the identity function, reducing training time and sometimes improving accuracy~\cite{antoran2020depth}.

\par
Data augmentation methods incorporate randomness to enlarge datasets and improve model robustness. In image tasks, augmentation can involve transformations such as rotation, cropping, flipping, color adjustments, kernel filtering, image mixing, random erasing, or neural style transfer. Some approaches extend augmentation to evaluation through test-time augmentation, introducing variability at inference to improve resilience~\cite{maleki2022generalizability}. These methods are used in algorithms such as Expectation--Maximization, posterior sampling, and Markov chain Monte Carlo methods~\cite{mumuni2022data}. Bootstrapping is another randomness-based technique, generating multiple resampled datasets through sampling with replacement, useful in ensemble learning to enhance stability and accuracy~\cite{tsamardinos2018bootstrapping}.

\par
Randomness is also a key factor in many advanced ML paradigms: Bayesian neural networks~\cite{magris2023bayesian}, variational autoencoders~\cite{wei2020recent}, reinforcement learning~\cite{ladosz2022exploration}, and even gradient noise injection~\cite{xiao2024noise}. Recent research has also examined PRNG usage in ML in relation to hardware performance and energy consumption~\cite{liu2020survey}. Kim et al.~\cite{kim2016dynamic}, for example, applied stochastic computing (SC) to deep neural networks, improving latency and power efficiency; SC, originally introduced by John von Neumann in the 1960s, encodes and processes information through random bitstreams. Liu et al.~\cite{liu2018energy}, however, noted that SC can be energy-inefficient for some deep learning applications.

\par
Transformer architectures---now central in domains from computer vision to natural language processing---also depend on randomness during training, including in SGD and dropout phases. Large language models such as GPT still rely on strong random generators for parameter initialization and regularization. The quality of these generators can influence the final system, as shown by Pranav et al.~\cite{dahiya2024machine}, who explored how attackers might exploit weaknesses in PRNGs to compromise ML systems. PRNGs also play a role in computational learning theory, including in criteria for Probably Approximately Correct (PAC) learning~\cite{daniely2021local}.

\par
Real-world applications further highlight PRNG importance: in microfluidic device studies of drop coalescence, random forest models have been applied~\cite{hu2024explainable,zhu2023analyzing}, relying on randomness for tree construction. Gundersen et al.~\cite{gundersen2022sources} list the lack of control over PRNG behavior as one source of irreproducibility in ML.

\par
In high-performance computing (HPC) and scientific simulations, PRNGs remain the standard, primarily because reproducibility is essential for debugging and verification. True random numbers are often too slow to generate and too large to store for large-scale simulations, making them impractical for workloads that require retracing execution. Applications such as high-energy physics or nuclear medicine simulations may require up to $10^{12}$ random numbers for a single replicate, with thousands of replicates to achieve statistical precision. Even with the fastest available storage, saving such quantities of true random data is infeasible.

\par
Quasi-random numbers can be used in specific contexts such as numerical integration in finance, but without certain improvements they suffer from limitations in high dimensions~\cite{sobol2011construction}. Quantum computing, with its intrinsic physical randomness, holds promise for future large-scale stochastic simulations~\cite{feynman2018simulating, cluzel2019quantum}, but until such systems are widely available, high-quality PRNGs remain the most efficient and reliable choice. These generators produce streams deterministically, with the generator’s source code serving as the ultimate proof of correlation structure. When properly designed, statistical tests fail to detect any structure---hence their common description as ``random numbers.'' However, it remains possible that future statistical tests will reveal weaknesses in today’s best PRNGs.

\par
The default PRNG in Python and PyTorch is the Mersenne Twister (MT)~\cite{matsumoto1998mersenne}, while TensorFlow defaults to Philox (with Threefry from the same cryptographically inspired family also available)~\cite{salmon2011parallel}. NumPy offers several PRNG choices; its default is PCG~\cite{oneill2014pcg}, but MT and Philox are also supported.

\par
Philox, Threefry, and ARS were introduced by Salmon et al. at the 2011 Supercomputing Conference. They use cryptographic techniques similar to AES, offering strong statistical properties though at a cost in speed. PCG, created in 2014 by O’Neill, claims superior statistical quality. MT, developed in 1998 by Matsumoto and Nishimura and updated in 2002, is known for its long period but also for failing certain statistical tests. Despite its weaknesses, MT remains one of the most used PRNGs in stochastic simulations.

\par
Linear congruential generators (LCGs) are among the earliest PRNG designs, simple and fast but prone to statistical deficiencies. Their quality depends heavily on parameters. For example, in our BigCrush tests from the TestU01 suite, a poorly chosen LCG with modulus $2^{31}$, multiplier $65539$, and increment $0$ achieved a period of $2^{29}$ and failed $125$ of the $144$ Crush tests, indicating severe deficiencies. A moderately better LCG with modulus $2^{48}$, multiplier $25214903917$, and increment $11$ had a period of $2^{48}$ but still failed $21$ Crush tests. Even a well-parameterized LCG with modulus $2^{63}$, multiplier $9219741426499971445$, and increment $1$---often considered ``good'' for practical purposes---failed $5$ Crush and $7$ BigCrush tests. These results highlight the wide performance gap among LCGs and the importance of PRNG selection for statistical robustness.

\par
PRNG quality is assessed using statistical test suites. Knuth proposed early tests in ``The Art of Computer Programming''~\cite{knuth2014art}. Marsaglia’s Diehard tests expanded on this with $15$ statistical tests, later extended by Brown et al. in the Dieharder suite. The NIST Statistical Test Suite (STS)~\cite{rukhin2001statistical} is widely used in cryptographic contexts. The most comprehensive is TestU01~\cite{l2007testu01}, offering multiple levels of testing: SmallCrush, Crush, and BigCrush.

\par
In this work, we use the TestU01 BigCrush battery as the reference. PCG and Philox are considered resistant to BigCrush failures, while MT is known to fail two tests. To compare with lower-quality generators, we include the aforementioned LCGs with varying parameter quality, as well as the standard C random generator, which uses a linear feedback shift register whose complexity depends on the available state size.

\par
The aim of this paper is to evaluate the extent to which the statistical quality of the random numbers influences results in stochastic simulations and ML applications. By testing PRNGs ranging from high-quality modern generators to deliberately weaker designs, and by applying them to representative workloads from both domains, we seek to quantify the performance, accuracy, and reproducibility impacts of PRNG choice.

\section{Related Work}
\label{sec:relatedwork}

The influence of the quality and source of randomness on computational tasks has been investigated in both machine learning (ML) and stochastic simulations, though the body of literature remains comparatively sparse.

\par
In the context of neural networks, Huk~\cite{huk2021random} explored the relationship between PRNG quality and classification performance in convolutional neural networks (CNNs) and multilayer perceptrons (MLPs). By drawing $95\%$ confidence intervals for quality measurements across different PRNGs, they demonstrated that variations in PRNG choice can lead to measurable changes in model performance, as evidenced by non-overlapping confidence intervals. These results suggest that the PRNG algorithm may influence training quality sufficiently to warrant adjustments in the interpretation of evaluation metrics. Koivu et al.~\cite{koivu2022quality} further established a correlation between PRNG statistical quality and the performance of dropout regularization in neural networks. Their findings reinforce the idea that generator quality can propagate through stochastic methods, influencing model generalization.

\par
Several studies have examined PRNG effects in simulation-based domains. For example, in~\cite{click2011quality} the authors compared three generators---the standard linear congruential generator (LCG), a modified LCG used in BOSS software, and the Mersenne Twister (MT)---for Monte Carlo simulations of liquid butane, methanol, and hydrated alanine polypeptides. While MT and the modified LCG produced similar results, the standard LCG yielded significant deviations, including up to $24\%$ higher average molecular volumes for methanol and up to $87\%$ larger volumes for hydrated tridecaalanine. These results highlight the potential for poor-quality PRNGs to introduce systematic biases in physical simulations.

\par
Beyond PRNG algorithm choice, several works have addressed the influence of random seeds on ML training outcomes. The study~\cite{picard2021torch} scanned up to $10^{4}$ seeds for popular computer vision architectures on CIFAR-10 and tested fewer seeds on ImageNet, revealing that while variance is generally small, extreme outlier seeds can produce significantly better or worse results than the mean. Similarly,~\cite{madhyastha2019model} quantified instability introduced by seed variation, finding that randomness can affect interpretability methods (e.g., attention maps, gradient-based explanations, LIME) and proposing Aggressive Stochastic Weight Averaging (ASWA) to reduce performance variance by $72\%$.

\par
The comparison between pseudo and true (or quantum) randomness has also attracted attention. Lebedev et al.~\cite{lebedev2024effects} showed that quantum random number generators (QRNGs) can yield statistically significant accuracy improvements over PRNGs for certain simulations, including approximating $\pi$ and Buffon’s needle, with potential error reductions up to $1.89\times$. Similarly,~\cite{bird2019effects} studied QRNG versus PRNG usage in initial weight distributions of dense and convolutional neural networks, as well as in decision tree splits. While QRNG occasionally outperformed PRNG in classification accuracy (e.g., $+2.82\%$ for EEG classification in dense networks), differences were often small and dataset-dependent.

\par
A broader ML-focused study,~\cite{jakob2022pitfalls}, assessed PRNG period length and determinism across multiple algorithms, finding that period length could significantly affect logistic regression, random forests, and LSTMs, while having minimal impact on linear regression. Likewise,~\cite{koivu2022quality} investigated five PRNGs for dropout in neural networks across four classification tasks, showing that true randomness could improve or degrade performance depending on the dataset and prediction problem.

\par
Finally, in the domain of large language models,~\cite{zhou2025assessing} assessed that seed choice can affect both macro-level metrics (accuracy, F1) and micro-level prediction consistency on GLUE and SuperGLUE benchmarks. Variance was significant enough to warrant explicit consideration of seed selection in LLM fine-tuning and evaluation pipelines.

\par
Overall, these works collectively show that PRNG quality, seed choice, and entropy source can all influence the performance, stability, and reproducibility of ML models and stochastic simulations. However, most studies have been limited to compare PRNGs to QRNGs, while we have seen earlier that QRNGs cannot be used in practice for large scale simulation. Studies also focus on specific architectures or simulation types, and few have systematically compared PRNGs of varying statistical quality across both domains. This gap motivates our work, which aims to quantify the extent to which PRNG quality affects results in representative ML and simulation workloads.

\section{Materials and methods}
\label{sec:materials}

The primary objective of this study was to evaluate whether the statistical quality of PRNGs has a measurable impact on the outcomes of stochastic applications. We focused primarily on PRNGs that are widely used in both ML and stochastic simulation, chosen for their strong statistical properties, but also included deliberately weaker generators to serve as baselines for comparison.

\par
To cover a range of computational domains and stochastic behaviors, we selected four representative applications. The first was a large-scale epidemiological agent-based model (ABM), from prior work by~\cite{hill2022reproductibilite}. This ABM reflects the complexity of HPC simulations, with numerous interacting agents and a high degree of stochasticity. TThe second and third applications were two independent implementations of the MNIST handwritten digit classification task, developed entirely from scratch: one using Python with NumPy\footnote{\url{https://github.com/yawen-d/Neural-Network-on-MNIST-with-NumPy-from-Scratch/tree/master}}, and another in~C++\footnote{\url{https://github.com/JanPokorny/mnist-from-scratch/tree/master}}. The fourth application was a RL environment implementing the CartPole task, also developed from scratch in Python\footnote{\url{https://github.com/Ancientkingg/cartpole}}. While the ABM is representative of large, high-dimensional simulations, the ML and RL cases serve as controlled, repeatable experiments where specific sources of randomness---such as weight initialization, batch selection, dropout, and environmental transitions---can be directly examined.

\par
For each application, the source code was modified to allow explicit selection of the PRNG. Seven different generators were tested. These included three linear congruential generators (LCGs) of progressively higher statistical quality:  
\begin{itemize}
    \item a ``bad'' LCG with parameters $m=2^{31}$, $a=65539$, $c=0$, known to fail the majority of TestU01 Crush tests (125 tests failed);
    \item a ``mid'' LCG with parameters $m=2^{48}$, $a=25214903917$, $c=11$, failing $21$ Crush tests;
    \item a ``good'' LCG with parameters $m=2^{63}$, $a=9219741426499971445$, $c=1$, which fails only a small number of BigCrush tests (7 tests failed).
\end{itemize}
In addition to these, we included the widely used MT, the PCG, the counter-based Philox generator, and the default C library \texttt{rand()} function, implemented as a linear-feedback shift register (LFSR)-based method. PCG and Philox are generally considered BigCrush-resistant, whereas MT is known to fail two BigCrush tests but remains a de facto standard in many scientific computing contexts.

\par
All experiments were conducted under a repeated-measures design. Each application was executed $30$ times for every PRNG, using fixed but distinct seeds to ensure reproducibility and to enable statistical analysis. This setup allowed computation of means and $95\%$ confidence intervals, as well as the application of parametric and non-parametric significance tests to detect differences between PRNGs. The full codebase, including PRNG-selection modifications, is available in the project’s public repository: \url{github double-blind}.

\par
Performance evaluation was tailored to each application. In the MNIST experiments, classification performance was quantified using accuracy, defined as the proportion of correctly classified samples in the validation set. The validation dataset was distinct from the training set to ensure that the reported accuracy reflected the model’s generalization capability. Accuracy was expressed as a percentage and aggregated across the $30$ runs for each PRNG to compute descriptive and inferential statistics.

\par
In the CartPole RL experiments, performance was measured by the average reward per episode, calculated as the sum of rewards obtained over all episodes divided by the number of episodes. Each time step in which the pole remained balanced yielded a reward of $+1$. Episodes terminated when the pole’s angle exceeded a specified threshold or the cart moved beyond the track boundaries. In this implementation, the maximum achievable reward per episode was $500$. Average reward thus reflected the agent’s sustained stability and control throughout an episode.

\par
For the epidemiological ABM, analysis focused on two critical epidemic indicators: the timing and amplitude of the infection peak. The maximum number of infections in the first epidemic peak was compared between PRNGs using one-way analysis of variance (ANOVA) under the assumptions of normality and homoscedasticity~\cite{miller1997beyond}. When normality was not met, the non-parametric Kruskal--Wallis test~\cite{kruskal1952use} was applied to the time step corresponding to the first infection peak to assess differences in timing. Beyond peak comparisons, entire epidemic time series were analyzed to assess similarity in temporal evolution.

\par
The choice of these specific applications allowed us to evaluate PRNG effects across scenarios differing in complexity, dimensionality, and stochastic dependency.

\section{Results}
\label{sec:results}

The epidemiological agent-based model exhibited a clear and statistically significant influence of the pseudorandom number generator on both the height and timing of infection peaks. Analysis of variance for peak height yielded $F=13.8692$ with $p=5.5162\times 10^{-15}$, corresponding to a large effect size of $\eta^{2}=0.2950$. Post-hoc Tukey HSD comparisons showed that the poor-quality LCG consistently produced peak values that were significantly different from all other PRNGs, whereas the remaining generators---mid- and good-quality LCGs, Mersenne Twister, PCG, Philox, and the C \texttt{rand()} implementation---were statistically indistinguishable from one another. 

\par
Similar results were observed for the timing of epidemic peaks. Here, the ANOVA returned $F=63.1213$ with $p=3.0604\times 10^{-50}$ and an even larger effect size ($\eta^{2}=0.6557$), again indicating that only the bad LCG produced timing patterns significantly different from the other generators.

\begin{figure}[htbp]
\centering
\includegraphics[width=1\linewidth]{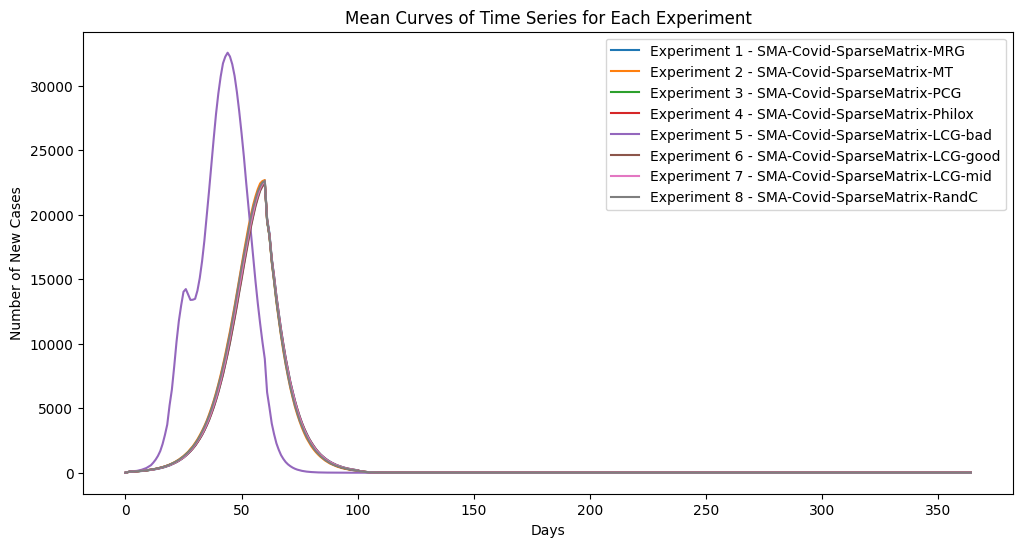}
\caption{Mean epidemic curves for each PRNG. The poor-quality LCG shows a visible displacement in amplitude and timing.}
\label{fig1}
\end{figure}

\par
The mean epidemic curves for each PRNG are presented in Figure~\ref{fig1}. With the exception of the poor-quality LCG, all curves overlap almost perfectly, indicating very similar epidemic dynamics. The trajectory obtained with the bad LCG is clearly displaced, with altered amplitude and peak timing, reflecting the statistical findings. 

\begin{figure}[htbp]
\centering
\includegraphics[width=1\linewidth]{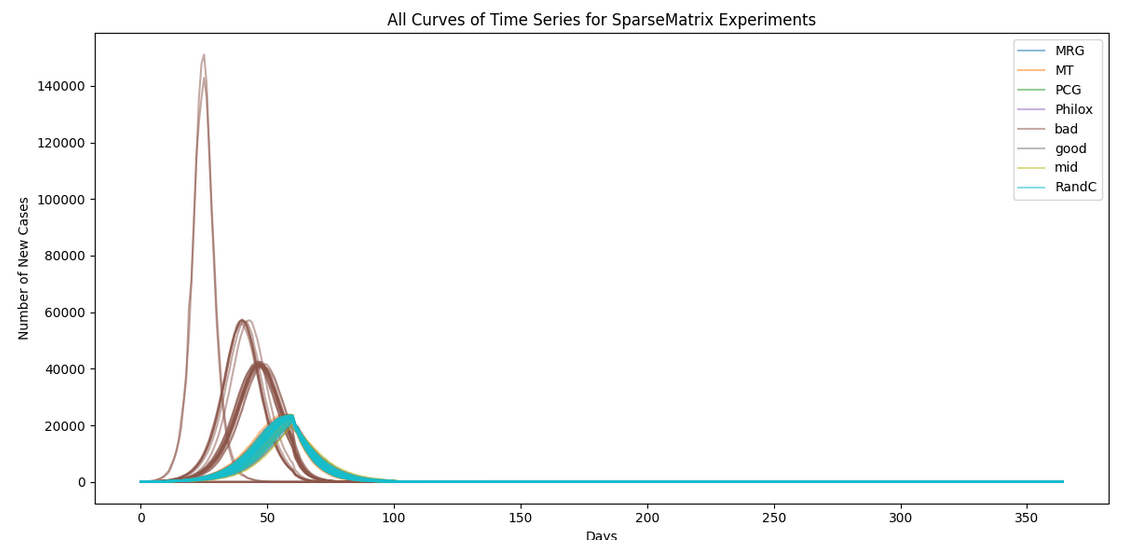}
\caption{Individual epidemic curves for each PRNG across $30$ replicates.}
\label{fig2}
\end{figure}

\par
Figure~\ref{fig2} shows all individual epidemic curves for all PRNGs across the $30$ replicates. For all high-quality generators, the curves remain tightly grouped despite stochastic variability, whereas the bad LCG yields several realizations with markedly different temporal patterns. These deviations, if occurring in real-world decision-support simulations, would represent substantial biases in both timing and severity projections.

\par
In the NumPy-based MNIST classification experiment, the choice of PRNG had a strong impact on accuracy. The ANOVA indicated $F=181.3880$, $p=1.0430\times 10^{-78}$, with a very large effect size ($\eta^{2}=0.8428$). All high-quality PRNGs achieved accuracies around $97.4$--$97.5\%$, whereas the bad LCG averaged $94.42\%$, a difference confirmed by the Tukey HSD analysis to be significant at $p<0.001$. No significant differences were found among the high-quality generators. 

\begin{figure}[htbp]
\centering
\includegraphics[width=1\linewidth]{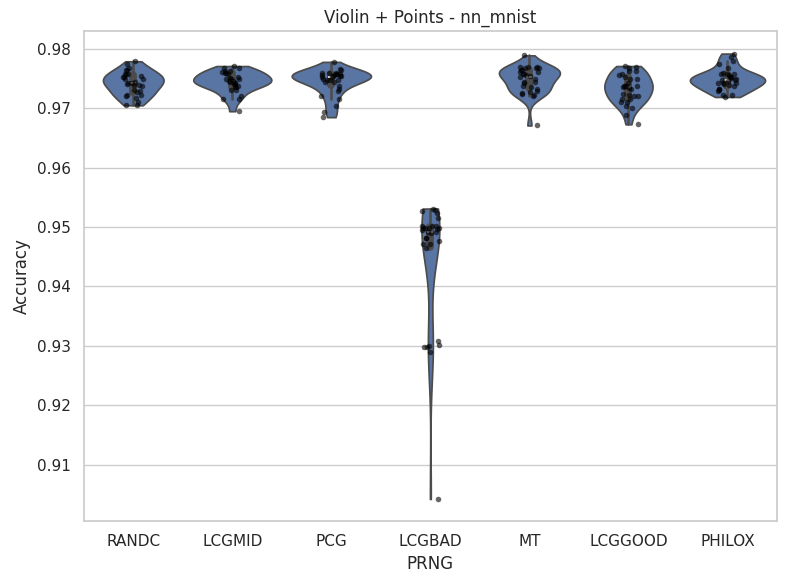}
\caption{Violin plots for MNIST (NumPy) classification accuracy by PRNG.}
\label{fig3}
\end{figure}

\begin{figure}[htbp]
\centering
\includegraphics[width=1\linewidth]{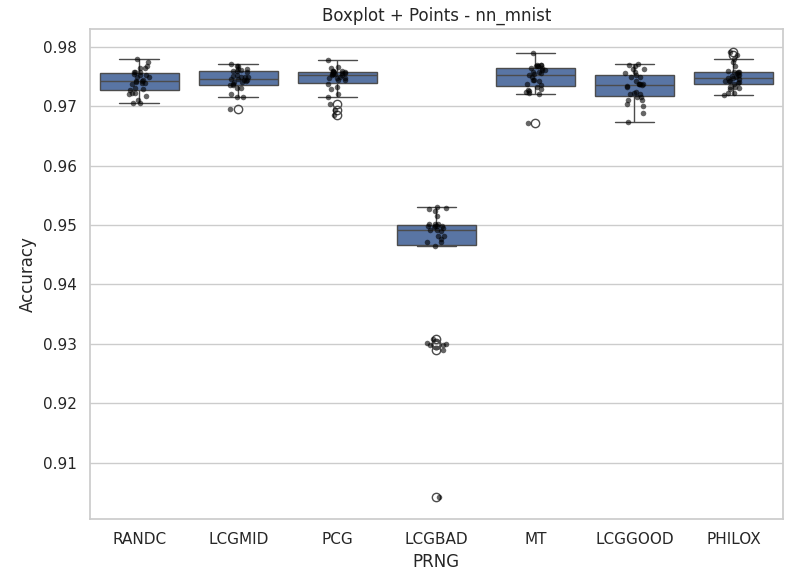}
\caption{Box plots for MNIST (NumPy) classification accuracy by PRNG.}
\label{fig4}
\end{figure}

\par
Figures~\ref{fig3} and~\ref{fig4} illustrate these results. The high-quality generators exhibit tight and nearly identical distributions, while the bad LCG’s distribution is both shifted downward and more dispersed. Non-parametric Kruskal--Wallis testing confirmed the same pattern ($p=7.3455\times 10^{-16}$), reinforcing the robustness of the finding.

\par
In contrast, the C++ MNIST implementation produced no statistically significant differences between PRNGs. The mean accuracies ranged from $79.47\%$ to $86.68\%$, with broad overlap of confidence intervals and an ANOVA result of $F=1.6162$, $p=0.144$, effect size $\eta^{2}=0.0456$. Standard deviations were notably higher for the bad LCG, Philox, and RandC, suggesting occasional unstable runs, but these differences did not reach statistical significance. 

\begin{figure}[htbp]
\centering
\includegraphics[width=1\linewidth]{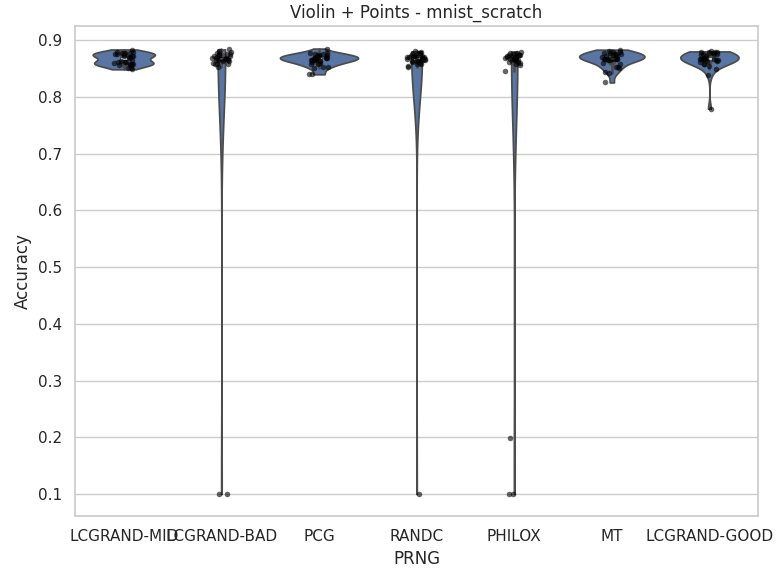}
\caption{Violin plots for MNIST (C++) classification accuracy by PRNG.}
\label{fig5}
\end{figure}

\begin{figure}[htbp]
\centering
\includegraphics[width=1\linewidth]{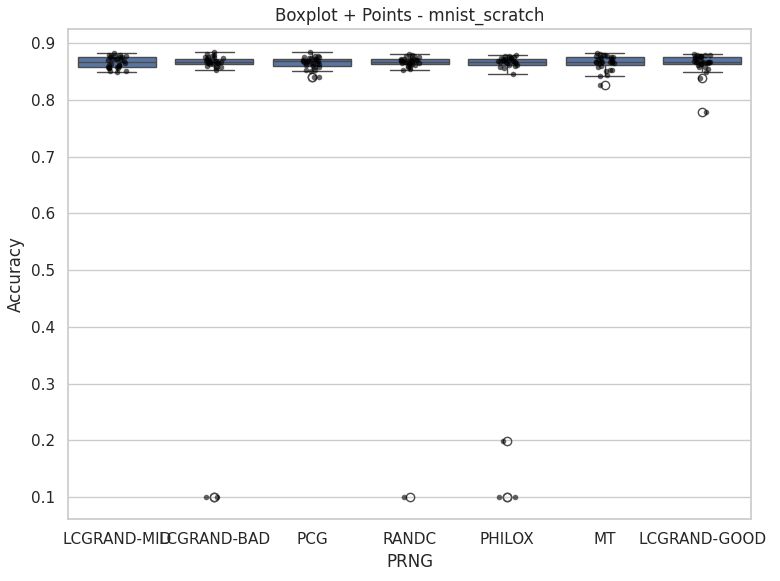}
\caption{Box plots for MNIST (C++) classification accuracy by PRNG.}
\label{fig6}
\end{figure}

\par
Figures~\ref{fig5} and~\ref{fig6} show the distributions for all generators, revealing generally similar central tendencies but greater variability for certain PRNGs. Non-parametric tests corroborated the absence of significant differences.

\par
The reinforcement learning CartPole experiment revealed the highest sensitivity to PRNG quality. The ANOVA indicated $F=149.7456$, $p=1.5386\times 10^{-71}$ with a large effect size ($\eta^{2}=0.8164$). The bad LCG produced an average reward of only $9.37$, indicating almost immediate failure in balancing the pole. Mid- and good-quality LCGs achieved moderate performance around $72$, while the top-performing generators---Mersenne Twister, PCG, Philox, and RandC---consistently produced average rewards between $285$ and $328$. Tukey HSD testing confirmed that all LCG variants differed significantly from the high-quality PRNGs, while the differences among the latter group were not statistically significant.

\begin{figure}[htbp]
\centering
\includegraphics[width=1\linewidth]{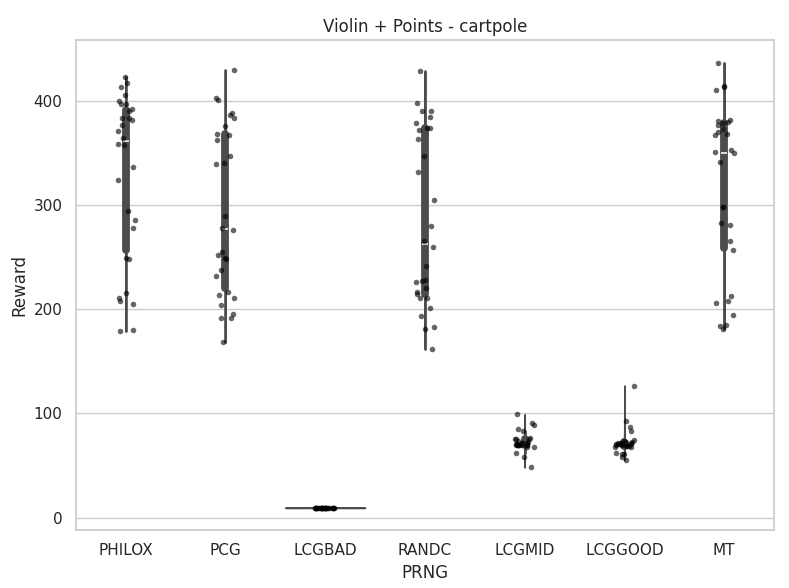}
\caption{Violin plots for average CartPole rewards by PRNG.}
\label{fig7}
\end{figure}

\begin{figure}[htbp]
\centering
\includegraphics[width=1\linewidth]{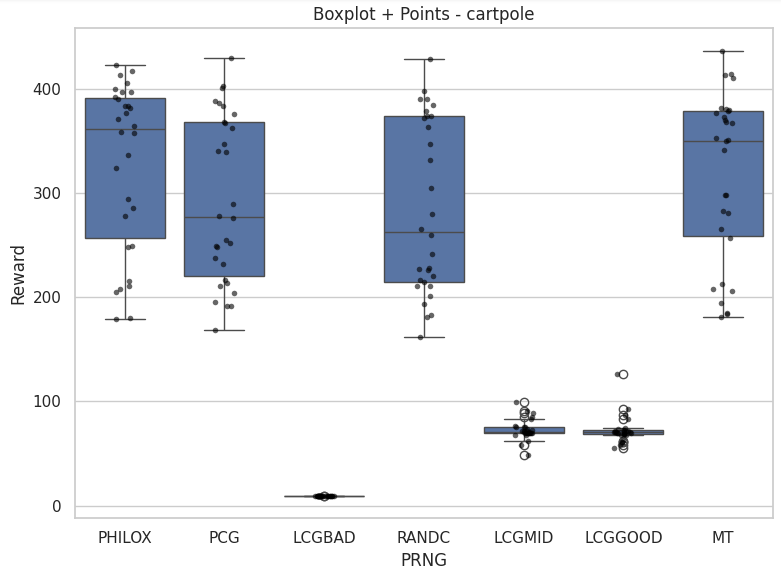}
\caption{Box plots for average CartPole rewards by PRNG.}
\label{fig8}
\end{figure}

\par
Figures~\ref{fig7} and~\ref{fig8} present these results, showing a clear separation between poor-quality generators, intermediate-quality LCGs, and top-tier PRNGs.

\section{Discussion}
\label{sec:discussion}

The results obtained across the four experimental settings reveal that the statistical quality of the pseudorandom number generator can influence stochastic computations to markedly different degrees depending on the application. 

\par
In the epidemiological ABM and the NumPy-based MNIST experiment, poor statistical quality in the PRNG resulted in significant and consistent deviations from the results obtained with higher-quality generators. In the ABM, these deviations manifested as shifts in both the timing and amplitude of epidemic peaks, patterns that, in a real-world policy context, could lead to incorrect conclusions regarding intervention timing and resource allocation. In the NumPy MNIST case, the lower accuracy and greater variability associated with the bad LCG suggest that poor-quality randomness can hinder convergence to optimal solutions in stochastic gradient descent training.

\par
In contrast, the C++ MNIST implementation showed no statistically significant dependence on PRNG choice, even though larger variability was observed for some generators. This may reflect differences in the implementation, numerical precision, or the reduced sensitivity of the specific training procedure to random variations in initialization and data shuffling. 

\par
The reinforcement learning CartPole experiment, however, demonstrated to be sensitive to PRNG quality. In this setting, mid- and good-quality LCGs produced intermediate results, while high-quality PRNGs achieved markedly better control and stability, and the bad LCG failed catastrophically.

\par
These findings suggest that the primary determinant of impact is not the algorithmic family of the generator---whether congruential, Mersenne Twister, or counter-based---but rather its measured statistical quality in standardized test suites such as TestU01 BigCrush. In our experiments, once a PRNG passed a threshold of statistical robustness, its results were generally indistinguishable from other top-tier generators, even if the underlying algorithmic principles differed. Philox, PCG, and Mersenne Twister, despite their distinct architectures, produced equivalent results in all tasks.

\par
Nevertheless, the consequences of using a low-quality PRNG in scientific computing remain serious. The bad LCG used in this study, which fails $125$ Crush tests in TestU01, consistently produced aberrant results in both simulations and machine learning, sometimes introducing systematic biases. This confirms that while performance considerations may drive PRNG selection for many applications---such as preferring Philox for parallel simulations or PCG for sequential workloads---the statistical quality of the generator must first meet a minimum standard.

\par
Overall, these results demonstrate that in most stochastic simulations and machine learning tasks, high-quality PRNGs of different families can be used interchangeably without affecting outcomes, and the choice may therefore be guided by performance, ease of implementation, or parallelization requirements.

\section{Conclusion}
\label{sec:conclusion}

This study evaluated the influence of PRNG statistical quality on the outcomes of both stochastic simulations and machine learning tasks, encompassing a large-scale epidemiological ABM, two from-scratch MNIST training implementations, and a reinforcement learning CartPole environment. By systematically varying the generator across seven PRNGs of differing statistical quality and repeating each experiment $30$ times with fixed seeds, we were able to isolate the effect of generator choice from other sources of variability.

\par
The results demonstrate that extremely poor statistical quality, exemplified by a low-parameter LCG failing most of the TestU01 Crush tests, can substantially distort simulation outputs and degrade ML performance. Such effects were evident in all experimental domains, ranging from shifts in epidemic peak timing and amplitude in the ABM to reduced classification accuracy in MNIST and almost complete failure in CartPole. 

\par
Once the statistical quality exceeded a certain threshold, however, the performance differences between generators became negligible in most settings. Mid- and good-quality LCGs, despite some statistical weaknesses, performed comparably to high-quality generators such as PCG, Philox, and Mersenne Twister in the ABM and MNIST experiments, though CartPole remained sensitive to generator quality.

\par
These findings suggest that for most stochastic simulations and ML tasks, PRNG selection can be based on computational performance, implementation convenience, and hardware suitability, provided that the chosen generator meets a robust statistical quality standard. Above all, the use of low-quality PRNGs in scientific computing should be avoided, as their deficiencies can propagate into significant and systematic errors in model outputs.

\section*{Acknowledgements}

The author thanks the maintainers of the open-source software and libraries used in this study. This research did not receive any specific grant from funding agencies in the public, commercial, or not-for-profit sectors.

 \bibliographystyle{elsarticle-num} 
 
 \bibliography{cas-refs}


\end{document}